\documentclass{ws-ijmpa}
\usepackage[super,compress]{cite}
\usepackage{graphicx}
\begin{document}
\markboth{Yu. A. Sitenko}{Chiral effects in magnetized quantum spinor matter in particle and astroparticle physics}

%
\catchline{}{}{}{}{}
%

\title{Chiral effects in magnetized quantum spinor matter in particle and astroparticle
physics}

\author{Yu. A. Sitenko}

\address{Bogolyubov Institute for Theoretical Physics,\\
National Academy of Sciences of Ukraine,\\
14-b Metrologichna Str., 03143 Kyiv, Ukraine}

\maketitle

\begin{history}
\received{Day Month Year}
\revised{Day Month Year}
\end{history}

\begin{abstract}
                        
Quantum spinor matter in extremal conditions (high densities and temperatures, presence of strong magnetic fields) have drawn the attention of researchers in diverse areas of contemporary physics, ranging from cosmology, high-energy and astroparticle physics to condensed matter physics. We study an impact of the confining boundary conditions on the properties of physical systems with hot dense magnetized ultrarelativistic spinor matter and elucidate a significant role of boundaries for such systems.

\end{abstract}

\ccode{PACS numbers:11.10.Wx, 03.70.+k, 71.70.Di, 73.23.Ra, 12.39.Ba, 25.75.Ld}


\section{Introduction}	

Magnetic fields of the order of the QCD energy scale squared can be produced in the quark-gluon plasma created in relativistic 
heavy-ion collisions (as a result of electric currents from the colliding charged ions); see Ref.\citen{Khar}. Magnetic fields up to $10^{15}$ Gauss may exist in some compact stars (magnetars); see Refs.\citen{Gle,Tur}. Even stronger fields are generated in progenitor magnetars during the core collapse after the supernovae explosion; see Ref.\citen{Fer}. Several competing mechanisms were proposed, providing for a generation of very strong magnetic fields in the early universe;\cite{Gra} despite 
the difference in details, the consensus is that rather strong magnetic fields should have been generated, since this is required by the present-day observation of weak, but nonvanishing, intergalactic magnetic fields. 

In the case of the inverse magnetic length, as well as temperature and chemical potential, exceeding considerably the mass of a relativistic quantized spinor matter field, 
it has been shown in theory \cite{Met,Fuk} that 
persistent and nondissipative currents emerge in thermal equilibrium, resulting in a variety of chiral effects in hot dense 
magnetized matter; see review in Ref.\citen{Mir} and references therein. 

So far chiral effects were mostly considered in unbounded matter, which perhaps may be relevant for cosmological applications. 
For all other applications (to astroparticle and high-energy physics), an account has to be taken of the 
finiteness of physical systems, and the role of boundaries in chiral effects in bounded matter should be clearly exposed. In the case of the quantized 
electromagnetic matter field, a choice of boundary conditions is motivated by material properties of the bounding plates, and, for instance, 
the Casimir effect \cite{Cas} is different for different boundary conditions: it is attractive between 
ideal-metal plates (i.e. made of material with an infinitely large magnitude of the dielectric permittivity), as well as 
between plates made of material with an infinitely large magnitude of the magnetic permeability; meanwhile, it is repulsive 
between an ideal-metal plate and an infinitely permeable one, see, e.g., Ref.\citen{Bor}. Thus, even if the material of plates is 
unknown for some reasons, it can in principle be determined by measuring the Casimir force or other physical quantities.

Namely such a situation happens in the case of the quantized spinor matter field, when nothing can be said about the ``material'' 
of boundaries, other than to admit that this ``material'' is impenetrable for spinor matter. Although the concept of quasiparticle excitations confined inside bounded-material samples is quite familiar in the context of 
condensed-matter physics, a quest for boundary conditions ensuring the confinement was initiated in particle physics in the 
context of a model description of hadrons as bags containing quarks \cite{Bog, Cho1}. Motivations for a concrete form of the 
boundary condition may differ in detail, but the key point is that the boundary condition has to guarantee the vanishing of the 
vector current of quark matter across the boundary, see Ref.\citen{Joh}. However, from this point of view, the bag boundary conditions 
proposed in Refs.\citen{Bog, Cho1} are not the most general ones. It has been rather recently realized that the most general 
boundary condition ensuring the confinenent of relativistic quantized spinor matter within a simply connected boundary involves 
four arbitrary parameters, \cite{Bee,Wie} and the explicit form 
of such a condition has been given. \cite{Si1,Si2,Si3} To study an 
impact of the background magnetic field on confined matter, one has to choose the magnetic field configuration with respect to the 
boundary surface. The primary interest is to understand the effect of a boundary which is transverse to the magnetic field strength 
lines. Then the simplest geometry is that of a slab of finite thickness in the uniform magnetic field directed perpendicular. It should 
be noted that such a geometry can be realized in condensed matter physics by putting slices of Dirac or Weyl semimetals \cite{Liu,Vaf} in an 
external transverse magnetic field. Note also that the slab geometry is conventional in a setup for the Casimir effect. \cite{Cas}  

Basing on the formalism of quantum field theory at nonzero temperature and chemical potential, we consider chiral effects in physical systems with hot dense magnetized spinor matter and elucidate a significant role of boundaries for such systems.

\section{Thermal equilibrium for chiral spinor matter}

We start with the operator of the second-quantized spinor field in a static background,
\begin{equation}
\hat{\Psi}({\bf r},t)=\sum\limits_{E_\lambda>0}e^{-{\rm i}E_\lambda t}\left\langle {\bf r}|\lambda\right\rangle\hat{a}_\lambda
+\sum\limits_{E_\lambda<0}e^{-{\rm i}E_\lambda t}\left\langle {\bf r}|\lambda\right\rangle\hat{b}_\lambda^{\dagger},\label{eq1}
\end{equation}
where $\hat{a}_\lambda^{\dagger}$ and $\hat{a}_\lambda$ ($\hat{b}_\lambda^{\dagger}$ and $\hat{b}_\lambda$) are the spinor particle (antiparticle) creation and destruction operators satisfying  anticommutation relations,
\begin{equation}
\left[\hat{a}_\lambda,\,\hat{a}_{\lambda'}^{\dagger}\right]_+ = \left[\hat{b}_\lambda,\,\hat{b}_{\lambda'}^{\dagger}\right]_+
=\left\langle \lambda|\lambda'\right\rangle, \label{eq2}
\end{equation}
and $\left\langle {\bf r}|\lambda\right\rangle$ is the solution to the stationary Dirac equation,
\begin{equation}
H\left\langle {\bf r}|\lambda\right\rangle=E_\lambda\left\langle {\bf r}|\lambda\right\rangle,\label{eq3}
\end{equation}
$H$ is the Dirac Hamiltonian operator, $\lambda$ is the set of parameters (quantum numbers) specifying a one-particle state, 
$E_\lambda$ is the energy of the state; wave functions $\left\langle {\bf r}|\lambda\right\rangle$ satisfy the requirement of 
orthonormality 
\begin{equation}
\int\limits_{\Omega}{\rm d}^3r\left\langle \lambda|{\bf r}\right\rangle\left\langle {\bf r}|\lambda'\right\rangle
=\left\langle \lambda|\lambda'\right\rangle\label{eq4}
\end{equation}
and completeness
\begin{equation}
\sum\left\langle {\bf r}|\lambda\right\rangle\left\langle \lambda|{\bf r}'\right\rangle=I\delta({\bf r}-{\bf r}');\label{eq5}
\end{equation}
summation is over the whole set of states, and $\Omega$ is the quantization volume. 

Conventionally, the operators of dynamical variables (physical observables) in second-quantized theory are defined as bilinears 
of the fermion field operator (1). One can define the fermion number operator,
\begin{equation}
\hat{N}=\frac{1}{2}\int\limits_{\Omega}{\rm d}^3r(\hat{\Psi}^{\dagger}\hat{\Psi}-\hat{\Psi}^T\hat{\Psi}^{\dagger T}) 
= \sum\left[\hat{a}_\lambda^{\dagger}\hat{a}_\lambda-\hat{b}_\lambda^{\dagger}\hat{b}_\lambda
-\frac{1}{2}{\rm sgn}(E_\lambda)\right], \label{eq6}
\end{equation}
and the energy (temporal component of the energy-momentum vector) operator,
\begin{equation}
\hat{P}^0=\frac{1}{2}\int\limits_{\Omega}{\rm d}^3r(\hat{\Psi}^{\dagger}H\hat{\Psi}-\hat{\Psi}^TH^T\hat{\Psi}^{\dagger T})
= \sum|E_\lambda|\left(\hat{a}_\lambda^{\dagger}\hat{a}_\lambda+\hat{b}_\lambda^{\dagger}\hat{b}_\lambda
-\frac{1}{2}\right),\label{eq7}
\end{equation}
where superscript $T$ denotes a transposition and ${\rm sgn}(u)$ is the sign function [${\rm sgn}(\pm u)=\pm 1$ at $u>0$]. A physical observable which is in general nonconserved is presented by an operator in the form
\begin{equation}
\hat{U}({\bf r})=\frac{1}{2}\left(\hat{\Psi}^{\dagger}\Upsilon\hat{\Psi}-\hat{\Psi}^T\Upsilon^T\hat{\Psi}^{\dagger T}\right),\label{eqno8}
\end{equation}
where matrix-valued differetial operator $\Upsilon$ is in general noncommuting with the Hamiltonian operator, 
$[\Upsilon, H]_- \neq 0$. The average of such an observable over the grand canonical ensemble is defined as 
\begin{equation}
\left\langle \hat{U}({\bf r})\right\rangle_{T,\mu}=\frac{{\rm Sp} \, \hat{U}({\bf r}) \, {\rm exp}\left[-(\hat{P}^0
-\mu\hat{N})/T\right]}{{\rm Sp} \, {\rm exp}\left[-(\hat{P}^0-\mu\hat{N})/T\right]}, \label{eq9}
\end{equation}
where equilibrium temperature $T$ is defined in the units of the Boltzmann constant, chemical potential is denoted by $\mu$, and ${\rm Sp}$ denotes the trace or the sum over the expectation values in the Fock state basis created by operators in (2).

We are considering the quantized charged spinor field in the background of 
a static uniform magnetic field with strength $\mathbf{B}=\boldsymbol{\partial}\times\mathbf{A}$, where 
$\mathbf{A}$ is the vector potential of the magnetic field. Assuming that the magnetic field is strong (supercritical) and ultrarelativistic spinor matter 
is at high temperature and high density, 
\begin{equation}
|e\mathbf{B}|>>m^2, \quad T>>m, \quad |\mu|>>m \label{eq10}
\end{equation}
($e$ is the charge of the matter field and natural units $\hbar=c=1$ are used), we shall neglect the mass of the spinor matter field, putting $m=0$ in the following. Thus the Dirac Hamiltonian operator takes form 
\begin{equation}
H=-{\rm i}\gamma^0\boldsymbol{\gamma}\cdot(\boldsymbol{\partial}-{\rm i}e\mathbf{A}). \label{eq11}
\end{equation}
Owing to the presence of chiral symmetry, 
\begin{equation}
 [H, \gamma^5]_- = 0, \label{eq12}
\end{equation}
where $\gamma^5=-{\rm i}\gamma^0\gamma^1\gamma^2\gamma^3$ ($\gamma^0$, $\gamma^1$, $\gamma^2$, and $\gamma^3$ are the generating 
elements of the Dirac-Clifford algebra, and $\gamma^5$ is defined according to Ref.\citen{Okun}, one can introduce also the 
following average:
\begin{equation}
\left\langle \hat{U}({\bf r})\right\rangle_{T,{\mu}_5}=\frac{{\rm Sp} \, \hat{U}({\bf r}) \, {\rm exp}\left[-(\hat{P}^0
-{\mu}_5\hat{N}^{5})/T\right]}{{\rm Sp} \, {\rm exp}\left[-(\hat{P}^0-{\mu}_5\hat{N}^{5})/T\right]}, \label{eq13}
\end{equation}
where 
\begin{equation}
\hat{N}^{5}=\frac{1}{2}\int\limits_{\Omega}{\rm d}^3r(\hat{\Psi}^{\dagger}\gamma^5\hat{\Psi}-
\hat{\Psi}^T\gamma^{5T}\hat{\Psi}^{\dagger T}) \label{eq14}
\end{equation}
is the axial charge and ${\mu}_5$ is the axial chemical potential. Diagonalizing simultaneously $\gamma^5$ and $H$, let us define their mutual complete system of wave functions $\left\langle {\bf r},\pm | \lambda \right\rangle$:
\begin{equation}
\gamma^5 \left\langle {\bf r},\pm |\, \lambda \right\rangle = \pm \left\langle {\bf r},\pm |\, \lambda \right\rangle \label{eq15} 
\end{equation}
and
\begin{equation}
H \left\langle {\bf r},\pm |\, \lambda \right\rangle = E_\lambda \left\langle {\bf r},\pm |\, \lambda \right\rangle. \label{eq16}
\end{equation}
Define further spectral densities which are relevant for the case of the observable given by 
$\hat{U}({\bf r})$:
\begin{equation}
\tau_{{\bf r},\pm}(E)={\rm tr}\Upsilon\left\langle {\bf r},\pm | \, \delta(H-E I) \, | {\bf r},\pm \right\rangle, \label{eq17}
\end{equation}
where ${\rm tr}$ denotes the trace over spinor indices.
Then averages (9) and (13) can be reduced to the form (see, e.g., Ref.\citen{SiGor})
\begin{equation}
\left\langle \hat{U}({\bf r})\right\rangle_{T,{\mu}}=-\frac{1}{2}\int\limits_{-\infty}^{\infty}dE\,\left[\tau_{{\bf r},+}(E) +\tau_{{\bf r},-}(E)\right]\,\tanh [(E-\mu)(2T)^{-1}] \label{eq18}
\end{equation}
and
\begin{eqnarray}
\left\langle \hat{U}({\bf r})\right\rangle_{T,{\mu}_5}=-\frac{1}{2}\int\limits_{-\infty}^{\infty}dE\,\left\{\tau_{{\bf r},+}(E)\,\tanh [(E-\mu_5)(2T)^{-1}] \right. \\ \nonumber
\left.+ \tau_{{\bf r},-}(E)\,\tanh [(E+\mu_5)(2T)^{-1}]\right\}. \label{eq19}
\end{eqnarray}

\section{Chiral effects in the unnbounded space}

A solution to the Dirac equation in the background of a static uniform magnetic field is well described in the literature, 
see, e.g., Ref.\citen{Akhie}. The one-particle energy spectrum in the case of the massless spinor field is
\begin{eqnarray}
E_{nk}=\left\{\begin{array}{l} \sqrt{2n|eB|+k^{2}}\\ [6 mm]
- \sqrt{2n|eB|+k^{2}}\end{array} \right\},
\quad -\infty<k<\infty, \quad n=0,1,2,...\, ,\label{20}
\end{eqnarray}
$k$ is the value of the wave number vector along the magnetic field, and $n$ enumerates the Landau levels. Unlike the lowest ($n=0$) Landau level, the Landau levels with $n \geq 1$ are doubly degenerate (in addition to the degeneracy owing to chirality): $\left\langle {\bf r},\pm, 1 |\, n, q, k \right\rangle$ and $\left\langle {\bf r},\pm, 2 |\, n, q, k \right\rangle$. Among the whole variety of spectral densities corresponding to different components of the vector and axial currents, only the spectral densities corresponding to components along the magnetic field are nonvanishing. Moreover, the levels with $n \geq 1$ do not contribute due to the cancellation between the '1' and '2' modes, and only the level with $n=0$ contributes. Choosing gauge $\mathbf{A}=(-yB, 0, 0,)$ with the magnetic field along the $z$-axis, $\mathbf{B}=(0, 0, B)$, and using the chiral representation for the Dirac matrices, 
$$
\gamma^0=\begin{pmatrix}
0\,&\,I\\
I\,&\,0
\end{pmatrix},\qquad
\boldsymbol{\gamma}=\begin{pmatrix}
0\,&-\boldsymbol{\sigma}\\
\boldsymbol{\sigma}&\,0
\end{pmatrix}, \quad 
\gamma^5=\begin{pmatrix}
-I\,&\,0\\
0\,&\,I
\end{pmatrix}
$$
($\sigma^1$, $\sigma^2$, and $\sigma^3$ are the Pauli matrices), one obtains the following expression for the modes corresponding to the lowest Landau level in the case of $eB>0$:
\begin{eqnarray}
\left\langle {\bf r},+ |\, 0, q, k \right\rangle |_{E_{0 k}>0} =\frac{1}{2\pi}{\rm e}^{{\rm i}(qx+kz)}\left(\frac{eB}{\pi}\right)^{1/4}{\rm exp}\left[-\frac{eB}{2}\left(y+\frac{q}{eB}\right)^2\right] \nonumber  \\
\times C_0\Biggl(0, \, 0, \, \Theta(-k), \, 0\Biggr)^T,
\label{21}
\end{eqnarray}
\begin{eqnarray}
\left\langle {\bf r},- |\, 0, q, k \right\rangle |_{E_{0 k}>0}=\frac{1}{2\pi}{\rm e}^{{\rm i}(qx+kz)}\left(\frac{eB}{\pi}\right)^{1/4}{\rm exp}\left[-\frac{eB}{2}\left(y+\frac{q}{eB}\right)^2\right] \nonumber  \\
\times \tilde{C}_0\Biggl(\Theta(k), \, 0, \, 0, \, 0\Biggr)^T,
\label{22}
\end{eqnarray}
\begin{eqnarray}
\left\langle {\bf r},+ |\, 0, q, k \right\rangle |_{E_{0 k}<0}=\frac{1}{2\pi}{\rm e}^{-{\rm i}(qx+kz)}\left(\frac{eB}{\pi}\right)^{1/4}{\rm exp}\left[-\frac{eB}{2}\left(y-\frac{q}{eB}\right)^2\right] \nonumber  \\
\times C_0\Biggl(0, \, 0, \, -\Theta(-k), \, 0\Biggr)^T
\label{23}
\end{eqnarray}
and
\begin{eqnarray}
\left\langle {\bf r},- |\, 0, q, k \right\rangle |_{E_{0 k}<0}=\frac{1}{2\pi}{\rm e}^{{-\rm i}(qx+kz)}\left(\frac{eB}{\pi}\right)^{1/4}{\rm exp}\left[-\frac{eB}{2}\left(y-\frac{q}{eB}\right)^2\right] \nonumber  \\
\times \tilde{C}_0\Biggl(\Theta(k), \, 0, \, 0, \, 0\Biggr)^T, \label{24}
\end{eqnarray}
where $\Theta(u)= \frac12 [1+{\rm sgn}(u)]$ is the step function, $-\infty<q<\infty$, and $|C_0|=|\tilde{C}_0|=1$; the modes in the case of $eB<0$ are obtained by charge conjugation, i.e. changing $eB\rightarrow -eB$ and multiplying the complex conjugates of the above modes by $\rm i\gamma^2$ (the energy sign is reversed). With the use of the explicit form of the modes, the nonanishing spectral densities are  immediately calculated:
\begin{equation}
\tau_{\pm}^z(E) \equiv {\rm tr}\,\gamma^0\gamma^z\left\langle {\bf r},\pm | \, \delta(H-E I) \, | {\bf r},\pm \right\rangle= \mp \frac{eB}{(2\pi)^2} \Theta(\mp E) \label{eq25}
\end{equation}
and
\begin{equation}
\tau_{\pm}^{z5}(E) \equiv {\rm tr}\,\gamma^0\gamma^z\gamma^5\left\langle {\bf r},\pm | \, \delta(H-E I) \, | {\bf r},\pm \right\rangle= - \frac{eB}{(2\pi)^2} \Theta(\mp E). \label{eq26}
\end{equation}
Determining averages according to (18) and (19),
\begin{equation}
\left\langle J^z\right\rangle_{T,{\mu}}=-\frac{1}{2}\int\limits_{-\infty}^{\infty}dE\,\left[\tau_{+}^z(E) +\tau_{-}^z(E)\right]\,\tanh [(E-\mu)(2T)^{-1}], 
\label{eq27}
\end{equation}
\begin{equation}
\left\langle J^{z5}\right\rangle_{T,{\mu}}=-\frac{1}{2}\int\limits_{-\infty}^{\infty}dE\,\left[\tau_{+}^{z5}(E) +\tau_{-}^{z5}(E)\right]\,\tanh [(E-\mu)(2T)^{-1}], \label{eq28}
\end{equation}
\begin{equation}
\left\langle J^z\right\rangle_{T,{\mu_5}}=-\frac{1}{2}\int\limits_{-\infty}^{\infty}dE\,\left\{\tau_{+}^z(E)\,\tanh [(E-\mu_5)(2T)^{-1}] + \tau_{-}^z(E)\,\tanh [(E+\mu_5)(2T)^{-1}]\right\} \label{eq29}
\end{equation}
and
\begin{equation}
\left\langle J^{z5}\right\rangle_{T,{\mu_5}}=-\frac{1}{2}\int\limits_{-\infty}^{\infty}dE\,\left\{\tau_{+}^{z5}(E)\,\tanh [(E-\mu_5)(2T)^{-1}] + \tau_{-}^{z5}(E)\,\tanh [(E+\mu_5)(2T)^{-1}]\right\}, \label{eq30}
\end{equation}
we obtain
\begin{equation}
\left\langle J^z\right\rangle_{T,{\mu_5}}=
-\frac{e B}{2\pi^2}{\mu}_5,   \label{eq31}
\end{equation}
\begin{equation}
\left\langle J^{z5}\right\rangle_{T,{\mu}} = 
-\frac{e B}{2\pi^2}\mu. \label{eq32}
\end{equation}
and
\begin{equation}
\left\langle J^{z}\right\rangle_{T,{\mu}} = \left\langle J^{z5}\right\rangle_{T,{\mu}_5} = 0. \label{eq33}
\end{equation}
Namely relations (31) and (32) are known as the chiral magnetic \cite{Fuk} and chiral separation \cite{Met} effects, respectively. Defining the left and right currents, 
\begin{equation}
\left\langle J^{zL}\right\rangle_{T,{\mu}_L} = \frac12 \left(\left\langle J^z\right\rangle_{T,{\mu_5}}+\left\langle J^{z5}\right\rangle_{T,{\mu}}\right)  \label{eq34}
\end{equation}
and
\begin{equation}
\left\langle J^{zR}\right\rangle_{T,{\mu}_R} = \frac12 \left(\left\langle J^z\right\rangle_{T,{\mu_5}}-\left\langle J^{z5}\right\rangle_{T,{\mu}}\right),  
\label{eq35}
\end{equation}
as well as the left and right chemical potentials, 
\begin{equation}
{\mu}_L = \frac12(\mu + {\mu}_5) \label{eq36}
\end{equation}
and
\begin{equation}
{\mu}_R = \frac12(\mu - {\mu}_5), \label{eq37}
\end{equation}
one can rewright (31) and (32) as
\begin{equation}
\left\langle J^{zL}\right\rangle_{T,{\mu}_L} =-\frac{e B}{2\pi^2}{\mu}_L  \label{eq38}
\end{equation}
and
\begin{equation}
\left\langle J^{zR}\right\rangle_{T,{\mu}_R} =\frac{e B}{2\pi^2}{\mu}_R.   \label{eq39}
\end{equation}
Actually, the chiral magnetic effect was first discovered in the form of (38) in Ref.\citen{Vil}.

\section{Chiral effects in a slab}

To study an influence of a background magnetic field on the properties of hot dense spinor matter, one has to account for the 
fact that the realistic physical systems are bounded. Our interest is in an effect of the static magnetic field with strength lines 
which are orthogonal to a boundary. Then, as was already noted, the simplest geometry of a material sample is that of a straight 
slab in the uniform magnetic field directed perpendicular. The one-particle energy spectrum in this case is [cf. (20)]
\begin{equation}
E_{nl}=\left\{\begin{array}{l} \sqrt{2n|eB|+k_l^{2}}\\ [6 mm]
- \sqrt{2n|eB|+k_l^{2}}\end{array} \right\},
\quad k_l > 0,\,\,\,\,n=0,1,2,\ldots,\label{eq40}
\end{equation} 
where the values of the wave number vector along the magnetic field, $k_l$ $\,$ ($l$ is integer), are to be determined by the 
boundary condition.

The most general boundary condition ensuring the confinement of relativistic spinor matter within a simply connected boundary 
is, see Refs.\citen{Si1,Si2,Si3},
\begin{equation}
\left\{I-\gamma^0\left[{\rm e}^{{\rm i}\varphi\gamma^5}\cos\theta  + (\gamma^1\cos\varsigma + 
\gamma^2\sin\varsigma)\sin\theta\right]{\rm e}^{{\rm i}\tilde{\varphi}{\gamma}^0(\boldsymbol{\gamma}\cdot\mathbf{n})}\right\} \chi(\mathbf{r})\left.\right|_{\mathbf{r}\in \partial\Omega}=0, \label{eq41}
\end{equation}
where $\mathbf{n}$ is the unit normal to surface $\partial\Omega$ bounding spatial region $\Omega$ 
and $\chi(\mathbf{r})$ is the wave function of the confined spinor matter, $\mathbf{r} \in \Omega$; 
matrices $\gamma^1$ and $\gamma^2$ in (41) are chosen to obey condition
\begin{equation}
[\gamma^1,\,\boldsymbol{\gamma}\cdot\mathbf{n}]_+=
[\gamma^2,\,\boldsymbol{\gamma}\cdot\mathbf{n}]_+=[\gamma^1,\,\gamma^2]_+=0,\label{eq42}
\end{equation}
and the boundary parameters in (41) are chosen to vary as
\begin{equation}
-\frac{\pi}{2}<\varphi\leq\frac{\pi}{2}, \quad -\frac{\pi}{2}\leq\tilde{\varphi}<\frac{\pi}{2}, \quad
0\leq\theta<\pi, \quad 0\leq\varsigma<2\pi. \label{eq43}
\end{equation}
The MIT bag boundary condition, \cite{Joh}
\begin{equation}
(I+{\rm i}\boldsymbol{\gamma}\cdot\mathbf{n})\chi(\mathbf{r})\left.\right|_{\mathbf{r}\in \partial\Omega}=0, 
\label{eq44}
\end{equation}
is obtained from (41) at $\varphi=\theta=0$, $\tilde{\varphi}=-{\pi}/{2}$.

The boundary parameters in (41) can be interpreted as the self-adjoint extension parameters. The self-adjointness of the 
one-particle energy (Dirac Hamiltonian in the case of relativistic spinor matter) operator in first-quantized theory is required 
by general principles of comprehensibility and mathematical consistency; see, e.g., Ref.\citen{Bon}. 
To put it simply, a multiple action is well defined for a self-adjoint operator only, allowing for the construction of functions 
of the operator, such as resolvent, evolution, heat kernel and zeta-function operators, with further implications upon second 
quantization.

In the case of a disconnected boundary consisting of two simply connected components, 
$\partial\Omega=\partial\Omega^{(+)}\bigcup\partial\Omega^{(-)}$, there are in general eight boundary parameters: 
$\varphi_{+}$, $\tilde{\varphi}_{+}$, $\theta_+$, and $\varsigma_+$ corresponding to $\partial\Omega^{(+)}$; and 
$\varphi_{-}$, $\tilde{\varphi}_{-}$, $\theta_{-}$, and $\varsigma_-$ corresponding to $\partial\Omega^{(-)}$. If spatial region $\Omega$ has the 
form of a slab bounded by parallel planes, $\partial\Omega^{(+)}$ and $\partial\Omega^{(-)}$, separated by distance $a$, then the boundary condition takes form  
\begin{equation}
\left(I-K^{(\pm)}\right)
\chi(\mathbf{r})\left.\right|_{z=\pm a/2}=0,\label{eq45}
\end{equation}
where 
\begin{equation}
K^{(\pm)}=\gamma^0\left[{\rm e}^{{\rm i}\varphi_{\pm}\gamma^5}\cos\theta_{\pm} + (\gamma^1\cos\varsigma_{\pm} + 
\gamma^2\sin\varsigma_{\pm})\sin\theta_{\pm}\right]{\rm e}^{\pm{\rm i}\tilde{\varphi}_{\pm}\gamma^0\gamma^z}, \label{eq46}
\end{equation}
coordinates $\mathbf{r}=(x,\,y,\,z)$ are chosen in such a way that $x$ and $y$ are tangential to the boundary, while $z$ is 
normal to it, and the position of $\partial\Omega^{(\pm)}$ is identified with $z=\pm a/2$. The confinement of matter inside the 
slab means that the vector bilinear, $\chi^{\dag}(\mathbf{r})\gamma^0{\gamma}^z\chi(\mathbf{r})$, vanishes at the slab boundaries,
\begin{equation}
\chi^{\dag}(\mathbf{r})\gamma^0{\gamma}^z\chi(\mathbf{r})\left.\right|_{z=\pm a/2}=0,\label{eq47}
\end{equation}
and this is ensured by condition (45). As to the axial bilinear, 
$\chi^{\dag}(\mathbf{r})\gamma^0{\gamma}^z\gamma^5\chi(\mathbf{r})$, it vanishes at the slab boundaries,
\begin{equation}
\chi^{\dag}(\mathbf{r})\gamma^0{\gamma}^z\gamma^5\chi(\mathbf{r})\left.\right|_{z=\pm a/2}=0,\label{eq48}
\end{equation}
in the case of $\theta_+ = \theta_- = \pi/2$ only, that is due to relation 
\begin{equation}
[K^{(\pm)}\left.\right|_{\theta_{\pm} = \pi/2}, \gamma^5]_- = 0. \label{eq49}
\end{equation}
However, note that, as a massless spinor particle is reflected from an impenetrable boundary, its helicity is flipped. 
Since the chirality equals plus or minus the helicity, chiral symmetry has to be necessarily broken by the confining boundary 
condition. Thus the case of $\theta_+ = \theta_- = \pi/2$ is not acceptable on the physical grounds. Moreover, there is a symmetry 
with respect to rotations around a normal to the slab, and the cases differing by values of 
$\varsigma_+$ and $\varsigma_-$ are physically indistinguishable, since they are related by such a rotation. The only way to 
avoid the unphysical degeneracy of boundary conditions with different values of $\varsigma_+$ and $\varsigma_-$ is to fix 
$\theta_+=\theta_-=0$. Then $\chi^{\dag}(\mathbf{r})\gamma^0{\gamma}^z\gamma^5\chi(\mathbf{r})$ is nonvanishing at the slab boundaries, and the boundary condition takes form
\begin{equation}
\left\{I-\gamma^0\exp\left[{\rm i}\left(\varphi_\pm\gamma^5\pm\tilde{\varphi}_\pm\gamma^0\gamma^z\right)\right]\right\}
\chi(\mathbf{r})\left.\right|_{z=\pm a/2}=0. \label{eq50}
\end{equation}
Condition (50) determines the spectrum of the wave number vector in the $z$ direction, $k_l$. The requirement that this spectrum 
be real and unambiguous yields constraint (see Refs.\citen{Si2,Si3})
\begin{equation}
\varphi_+=\varphi_-=\varphi, \quad \tilde{\varphi}_+=\tilde{\varphi}_-=\tilde{\varphi}; \label{eq51}
\end{equation}
then the $k_l$ spectrum is determined implicitly from relation 
\begin{equation}
k_l\sin\tilde{\varphi}\cos(k_l a)+(E_{... l}\cos\tilde{\varphi}-m\cos\varphi)\sin(k_l a)=0, \label{eq52}
\end{equation}
where $m$ is the mass of the spinor matter field and $E_{... l}$ is the energy of the one-particle state.
In the case of the massless spinor matter field, $m=0$, and the background uniform magnetic field perpendicular to the slab,
${\bf B}=(0,0,B)$, $E_{... l}$ takes the form of $E_{n l}$ (40), and relation (52) is reduced to
\begin{equation}
k_l\sin\tilde{\varphi}\cos(k_l a)+E_{n l}\cos\tilde{\varphi}\sin(k_l a)=0, \label{eq53}
\end{equation}
depending on one parameter only, although the boundary condition depends on two parameters,
\begin{equation}
\left\{I-\gamma^0\exp\left[{\rm i}\left(\varphi\gamma^5 \pm  \tilde{\varphi}\gamma^0\gamma^z\right)\right]\right\}
\chi(\mathbf{r})|_{z=\pm a/2}=0. \label{eq54}
\end{equation}

A solution to the Dirac equation,
\begin{equation}
 H \left\langle {\bf r} |\, n, q, k_l \right\rangle = E_{n l} \left\langle {\bf r} |\, n, q, k_l \right\rangle, \label{eq55}
\end{equation}
satisfying boundary condition (54), is a standing wave inside a slab, which is composed from counterpropagating waves of opposite chiralities. For instance, we get for the lowest Landau level in the case of $e B > 0$:
\begin{eqnarray}
\left\langle {\bf r} |\, 0, q, k_l \right\rangle |_{E_{0 l}>0}=\frac{1}{2\pi}{\rm e}^{{\rm i}qx}\left(\frac{eB}{\pi}\right)^{1/4}{\rm exp}\left[-\frac{eB}{2}\left(y+\frac{q}{eB}\right)^2\right] \nonumber  \\
\times \Biggl(C_0{\rm e}^{{\rm i}k_lx}, \, 0, \tilde{C}_0{\rm e}^{-{\rm i}k_lx}, \, 0\Biggr)^T
\label{56}
\end{eqnarray}
and
\begin{eqnarray}
\left\langle {\bf r} |\, 0, q, k_l \right\rangle |_{E_{0 l}<0}=\frac{1}{2\pi}{\rm e}^{-{\rm i}qx}\left(\frac{eB}{\pi}\right)^{1/4}{\rm exp}\left[-\frac{eB}{2}\left(y-\frac{q}{eB}\right)^2\right] \nonumber  \\
\times \Biggl(\tilde{C}_0{\rm e}^{-{\rm i}k_lx}, \, 0, \, - C_0{\rm e}^{{\rm i}k_lx}, \, 0\Biggr)^T,
\label{57}
\end{eqnarray}
where $k_l=\left[l\pi - {\rm sgn}(E_{0 l})\tilde{\varphi}\right]/a$, $-\infty<q<\infty$, and $|C_0| = 
|\tilde{C}_0| = \sqrt{\pi/a}$.
Using the explicit form of standing waves, one can compute spectral densities corresponding to different components of the vector and axial currents. We obtain
\begin{equation}
\tau^{x}(E)=\tau^{y}(E)=\tau^{z}(E)=0 \label{58}
\end{equation}
and
\begin{equation}
\tau^{x5}(E)=\tau^{y5}(E)=0, \label{59}
\end{equation}
whereas only the $n=0$ level contributes to the only nonvanishing spectral density corresponding to the  axial current component along the magnetic field,
\begin{equation}
\tau^{z5}(E)=-\frac{eB}{2\pi a}\sum\limits_{k_l^{(\pm)}>0}\delta \left(\pm k_l^{(\pm)} - E\right) \Theta(\pm E), \quad   k_l^{(\pm)}=\left(l\pi \mp \tilde{\varphi}\right)/a; \label{eq60}
\end{equation}
the latter relation is valid both at $eB>0$ and at $eB<0$.
Consequently, the following expression is obtained:
\begin{eqnarray}
\left\langle J^{z5}\right\rangle_{T,{\mu}} \equiv -\frac{1}{2}\int\limits_{-\infty}^{\infty}dE\,\tau^{z5}(E)\,\tanh [(E-\mu)(2T)^{-1}]  \nonumber \\
=\frac{eB}{4\pi a} \sum\limits_{\pm} \sum\limits_{k_l^{(\pm)}>0}\tanh \left[\left(\pm k_l^{(\pm)} - \mu\right)(2T)^{-1}\right].  \label{eq61}
\end{eqnarray}
The use of relation which was proven in Ref.\citen{Si4},
\begin{eqnarray}
\sum\limits_{n \geq 0}\frac{y\sin x}{\cos x+\cosh[(2n+1)\pi y]} 
=\frac{1}{2\pi} \int\limits_{0}^{\infty}{\rm d}\eta \,\frac{\sin x\sinh(2\pi/y)}{(\cos x+{\rm cosh} \eta)
[{\rm cosh}(2\pi/y)+\cos(\eta/y)]} \nonumber \\
-\frac{{\rm sinh}\left\{2[{\rm arctan}\left({\rm tan}\frac{x}{2}\right)]/y\right\}}{{\rm cosh}(\pi/y)+{\rm cosh}\{2[{\rm arctan}\left(\tan\frac{x}{2}\right)]/y\}}, \label{eq62}
\end{eqnarray}
allows us to get the final form for the axial current:
\begin{eqnarray}
\left\langle J^{z5}\right\rangle_{T,{\mu}}=-\frac{eB}{2\pi a} \Biggl\{{\rm sgn}(\mu)F\Biggl(|\mu|a + {\rm sgn}(\mu)\left[\tilde{\varphi}-
{\rm sgn}(\tilde{\varphi}){\pi}/{2}\right];Ta\Biggr) \Biggr. \nonumber \\
\Biggl. - \frac{1}{\pi}\left[\tilde{\varphi}-
{\rm sgn}(\tilde{\varphi}){\pi}/{2}\right]\Biggr\} , \label{eq63}
\end{eqnarray}
where 
\begin{eqnarray}
F(s;t)=\frac{s}{\pi}
-\frac{1}{\pi}\int\limits_{0}^{\infty}{{\rm d}v\,\frac{\sin(2s){\rm sinh}(\pi/t)}
{[\cos(2s)+{\rm cosh}(2v)][{\rm cosh}(\pi/t)+\cos(v/t)]}}\nonumber \\ 
+\frac{{\rm sinh} \left\{[{\rm arctan} ({\rm tan} s)]/t\right\}}{{\rm cosh}[\pi/(2t)]+{\rm cosh}
\left\{[{\rm arctan} ({\rm tan} s)]/t\right\}}.\label{eq64}
\end{eqnarray}

In view of relation
\begin{equation}
\lim\limits_{a\rightarrow \infty}\frac{1}{a}F(|\mu|a;Ta)=|\mu|/{\pi},\label{eq65}
\end{equation}
the case of a magnetic field filling the whole (infinite) space \cite{Met} is obtained from (63) as a limiting case [cf. (32)],
\begin{equation}
\lim\limits_{a\rightarrow \infty}\left\langle J^{z5}\right\rangle_{T,{\mu}}=-\frac{eB}{2\pi^2} \mu.\label{eq66}
\end{equation}
Unlike this unrealistic case, the realistic case of a magnetic field confined to a slab of finite thickness is temperature dependent. In particular, we get
\begin{equation}
\left\langle J^{z5}\right\rangle_{0,{\mu}}=-\frac{eB}{2\pi a}  \Biggl[{\rm sgn}(\mu)\left[\!\!\left[\frac{|\mu|a+
{\rm sgn}(\mu)\tilde{\varphi}}{\pi} 
+ \Theta(-\mu\tilde{\varphi})\right]\!\!\right] - \frac{\tilde{\varphi}}{\pi} + 
\frac{1}{2}{\rm sgn}(\tilde{\varphi})\Biggr] \label{eq67}
\end{equation}
and
\begin{equation}
\left\langle J^{z5}\right\rangle_{\infty,{\mu}}=-\frac{eB}{2\pi^2} \mu, \label{eq68}
\end{equation} 
where $[\![u]\!]$ denotes the integer part of quantity $u$ (i.e. the integer which is less than or equal to $u$); see Fig.1., where the axial currents at $T=0$ and at $T=\infty$ are presented as functions of chemical potential for different values of boundary parameter $\tilde{\varphi}$ ($eB > 0$ is taken for definiteness). As follows from (63), the boundary condition can serve as a source which is additional to the spinor matter density: the contribution of the boundary to the axial current effectively enhances (at $-\pi/2 < \tilde{\varphi}<0$) or diminishes (at $0<\tilde{\varphi} < \pi/2$) the contribution of chemical potential; the situation is reversed in the case of $eB < 0$. Because of the boundary condition, the chiral separation effect can be nonvanishing even at zero chemical potential,
\begin{equation}
\left\langle J^{z5}\right\rangle_{T,0}=-\frac{eB}{2\pi a}\Biggl\{F(\tilde{\varphi}-{\rm sgn}(\tilde{\varphi})\pi/2; Ta) - \frac{1}{\pi}\left[\tilde{\varphi}-{\rm sgn}(\tilde{\varphi}){\pi}/{2}\right]\Biggr\}; \label{eq69}
\end{equation}
the latter vanishes in the limit of infinite temperature, 
\begin{equation}
\left\langle J^{z5}\right\rangle_{\infty,0}=0.\label{eq70}
\end{equation}

\begin{figure}[b]
\centerline{\includegraphics[width=10.8cm]{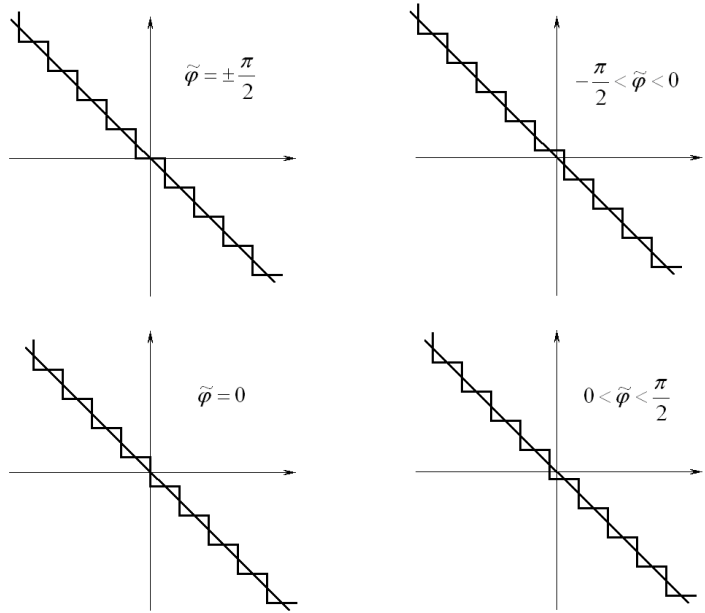}}
\caption{A stepwise behavior of the axial current at $T=0$ is smoothed out as $T \rightarrow \infty$. \label{f1}}
\end{figure}

\section{Conclusion}

We have considered chiral effects in hot dense magnetized quantum relativistic spinor matter. An issue of the confining boundary condition plays the key role in this survey. In the absence of boundaries there exist the chiral magnetic effect which is exhibited by the nondissipative vector current along the magnetic field, see (31), and the chiral separation effect which is exhibited by the nondissipative axial current in the same direction, see (32); both currents are temperature independent. As boundaries are introduced and the matter volume is shrinked to a slab which is transverse to the magnetic field, the fate of these currents is different. The axial current stays on, becoming dependent on temperature, on a thickness of the slab and on a choice of boundary conditions, see 
(62) and (63); as temperature increases from zero to large values, a stepped-shape behaviour of the axial current as a function of chemical potential is changed to a smooth one, see (66) and (67). The vector current is extinct, which is due to the vanishing of the appropriate spectral density, see (58). The average over the ensemble with the axial chemical potential cannot be introduced, because the confining boundary condition violates chiral symmetry: standing waves inside a slab are formed from counterpropagating waves of opposite chiralities, see e.g. (56) and (57). Thus, the chiral magnetic effect in a slab is eliminated by the confining boundary condition.

It should be recalled that, in the case of zero temperature and chemical potential, the pressure from the vacuum in a slab of confined charged massive matter in the background of magnetic field is repulsive and independent of the choice of a boundary condition, as well as of the slab thickness; see Refs.\citen{Si1,Si2,Si3}. Contrary to this, boundary conditions play a significant role in the case of nonzero temperature and chemical potential, and chiral effects in hot dense magnetized spinor matter confined to  a slab depend both on the choice of a boundary condition and on the slab thickness; moreover, the boundary condition can serve as a source that is additional to the spinor matter density.

\section*{Acknowledgments}

I would like to thank the Organizers of the SEENET-MTP  Workshop `Field Theory and the Early Universe' (BW2018) for kind hospitality during this interesting and inspiring meeting. The work was supported by the National Academy of Sciences of Ukraine (Project No.01172U000237),
by the Program of Fundamental Research of the Department of Physics and Astronomy of the National Academy of Sciences of Ukraine (Project No.0117U000240) and by the ICTP -- SEENET-MTP project NT-03
`Cosmology - Classical and Quantum Challenges'.


\end{document}